\documentclass[conference]{IEEEtran}

\usepackage{amsmath}
\usepackage{amsthm}
\usepackage{amssymb}
\usepackage{bm}
\usepackage{xspace}
\usepackage{xcolor}
\usepackage{graphicx}
\usepackage{url}
\usepackage{framed}
\usepackage{float}
\usepackage{rotating}
\usepackage{verbatim}
\usepackage{listings}
\usepackage{lscape}
\usepackage{cite}
\usepackage{pgfplots}
\usepackage{pgf}
\usepackage{tikz}
\usetikzlibrary{arrows,shapes.misc,chains,scopes}
\usepackage{pgfplotstable}
\usepackage{booktabs}
\usepackage{colortbl}

\newcommand{\executeiffilenewer}[3]{%
\ifnum\pdfstrcmp{\pdffilemoddate{#1}}%
{\pdffilemoddate{#2}}>0%
{\immediate\write18{#3}}\fi%
}

\graphicspath{{figures/}}

\theoremstyle{plain}

\newcounter{algocount}
\newcounter{examplecount}

\newenvironment{algorithm}[1][]{\refstepcounter{algocount}\setlength{\parindent}{0.5cm}\begin{trivlist}\item \textbf{Algorithm \thealgocount.}#1\\[-0.2cm]\rule{\columnwidth}{1pt}}{\\[-0.2cm]\rule{\columnwidth}{1pt}\end{trivlist}}

\newcommand{\veczero}{\boldsymbol{0}}

\newcommand{\vecp}{\boldsymbol{p}}
\newcommand{\vecq}{\boldsymbol{q}}
\newcommand{\vecr}{\boldsymbol{r}}

\newcommand{\vecw}{\boldsymbol{w}}
\newcommand{\vecx}{\boldsymbol{x}}

\newcommand{\mathh}{\boldsymbol{H}}

\newcommand{\capacity}{\ensuremath{\mathsf{C}}\xspace}

\newcommand{\setr}{\ensuremath{\mathbf{R}}\xspace}

\newcommand{\setrpp}{\ensuremath{\mathbf{R}_{>0}}\xspace}

\newcommand{\bpm}{\begin{pmatrix}}
\newcommand{\epm}{\end{pmatrix}}

\DeclareMathOperator*{\argmax}{argmax}

\DeclareMathOperator{\entop}{\mathbb{H}}
\DeclareMathOperator{\miop}{\mathbb{I}}

\DeclareMathOperator*{\maximize}{maximize}
\DeclareMathOperator*{\st}{subject\;to}

\usepackage{cite}
\newcommand{\bicmmi}{\miop^\mathrm{bicm}}

\title{An Efficient Algorithm to Calculate BICM Capacity}

\IEEEoverridecommandlockouts

\author{\IEEEauthorblockN{Georg B\"ocherer\IEEEauthorrefmark{1}, Fabian Altenbach\IEEEauthorrefmark{4}, Alex Alvarado\IEEEauthorrefmark{3}, Steven Corroy\IEEEauthorrefmark{4}, and Rudolf Mathar\IEEEauthorrefmark{4}}
\IEEEauthorblockA{\IEEEauthorrefmark{1}Institute for Communications Engineering, Technische Universit\"at M\"unchen, Germany\\
Email: \texttt{georg.boecherer@tum.de}}
\IEEEauthorblockA{\IEEEauthorrefmark{4}Institute for Theoretical Information
Technology, RWTH Aachen University, Germany\\
Email: \texttt{\{altenbach,corroy,mathar\}@ti.rwth-aachen.de}}
\IEEEauthorblockA{\IEEEauthorrefmark{3}Signal Processing and Communications Laboratory, University of Cambridge, UK\\
Email: \texttt{alex.alvarado@ieee.org}\\
}
\thanks{This work was supported by the German Ministry of Education and Research in the framework of an Alexander von Humboldt Professorship, by the UMIC Research Center, RWTH Aachen University, by The British Academy and The Royal Society (via the Newton International Fellowship scheme), UK, and by the European Community's Seventh's Framework Programme (FP7/2007-2013) under grant agreement No. 271986.}
}

\begin{document}

\maketitle
\begin{abstract}
Bit-interleaved coded modulation (BICM) is a practical approach for reliable communication over the AWGN channel in the bandwidth limited regime. For a signal point constellation with $2^m$ points, BICM labels the signal points with bit strings of length $m$ and then treats these $m$ bits separately both at the transmitter and the receiver. BICM capacity is defined as the maximum of a certain achievable rate. Maximization has to be done over the probability mass functions (pmf) of the bits. This is a non-convex optimization problem. So far, the optimal bit pmfs were determined via exhaustive search, which is of exponential complexity in $m$. In this work, an algorithm called \emph{bit-alternating convex concave method} (\textsc{Bacm}) is developed. This algorithm calculates BICM capacity with a complexity that scales approximately as $m^3$. The algorithm iteratively applies convex optimization techniques. \textsc{Bacm} is used to calculate BICM capacity of $4,8,16,32$, and $64$-PAM in AWGN. For PAM constellations with more than $8$ points, the presented values are the first results known in the literature.
\end{abstract}

\section{Introduction}

Bit-interleaved coded modulation (BICM) \cite{Zehavi92,Caire98,Fabregas2008} is a \emph{de facto} standard for wireless communications, and it is used in e.g., HSPA, IEEE 802.11a/g/n, and the latest DVB standards (DVB-T2/S2/C2).

In BICM, signal points from a finite constellation are labeled with bit strings. E.g., for $16$-PAM, the signal points are labeled with $\log_2 16=4$ bits each. The bits in the labels are then treated independently both at the transmitter and the receiver. According to \cite{Fabregas2010}, to determine BICM capacity, a certain achievable rate has to be maximized over the bit probability mass functions (pmf). We will make this statement precise later in this work. This maximization is a non-convex optimization problem \cite[Fig.~1]{Alvarado2011}. So far, BICM capacity has been calculated using exhaustive search only. For the AWGN channel, results are presented for $8$-PAM in \cite[Fig.~3]{Agrell10b} and \cite[Fig.~1]{Alvarado2011} and for $16$-QAM in \cite[Fig.~2]{Fabregas2010}. The complexity of exhaustive search is exponential in the number of bits in the labels, and calculating BICM capacity becomes an intractable problem for large constellations. This motivates the present work.

Our approach is as follows. We start by considering a discrete memoryless channel (DMC) operated by a BICM transceiver. To calculate BICM capacity, we develop a new algorithm called \emph{bit alternating convex-concave method} (\textsc{Bacm}), which combines two optimization techniques: first, maximization is done sequentially over one bit pmf at a time, and second, the maximization over one bit pmf is done using the convex-concave procedure \cite{Yuille2003}. We then show how an average power constraint can be taken into account by \textsc{Bacm}. This allows us to use \textsc{Bacm} to calculate BICM capacity of PAM constellations in AWGN. We provide numerical results for $4$ and $8$-PAM and, for the first time in the literature, for $16$, $32$, and $64$-PAM. The results show that BICM capacity is close to AWGN capacity and significantly larger than what can be achieved by operating BICM with uniform bit pmfs. Finally, we argue that the complexity of \textsc{Bacm} scales approximately as $m^3$  and logarithmically in the precision with which the optimal bit pmfs are calculated. An implementation of \textsc{Bacm} in Matlab is available on our website \cite{website:bacm}.

\section{System Model and Problem Statement}
\label{sec:problem}
Consider a DMC with $2^m$ input symbols $\mathcal{X}=\{1,\dotsc,2^m\}$ and $n$ output symbols $\mathcal{Y}=\{1,\dotsc,n\}$. The channel is specified by a matrix of transition probabilities $\mathh\in\setr^{n\times 2^m}$, where $\setr$ denotes the set of real numbers. The input of the channel is the random variable $X$, which takes values in $\mathcal{X}$ according to the pmf $\vecp$. The channel output is the random variable $Y$, which takes values in $\mathcal{Y}$ according to the pmf $\vecr=\mathh\vecp$.
\subsection{DMC Capacity}
We denote the mutual information between $X$ and $Y$ either by $\miop(X;Y)$ or by $\miop(\vecp)$. The DMC capacity is \cite[Eq. (7.1)]{Cover2006}
\begin{align}
\begin{split}
\capacity=\max_{\vecp}\quad&\miop(\vecp).
\end{split}
\end{align}
The maximization is a convex optimization problem \cite[Prob.~4.57]{Boyd2004} and it can be solved by the Blahut-Arimoto algorithm \cite{Blahut1972,Arimoto1972} or by a software package such as CVX \cite{CVX}.

\subsection{BICM Capacity}
In BICM, the input symbols are represented by their $m$-bit binary expansion, i.e,
\begin{align}
\begin{split}
1&\leftrightarrow\overbrace{\texttt{0}\dotsb\texttt{00}}^{m\text{ bits}}\\
2&\leftrightarrow\texttt{0}\dotsb\texttt{01}\\
&\;\;\,\vdots\\
2^m&\leftrightarrow\texttt{1}\dotsb\texttt{11}.
\end{split}\label{eq:labeling}
\end{align}
Each bit position of the channel input is treated independently both at the transmitter and the receiver, see 
\cite{Fabregas2008,Fabregas2010} for details. This leads to the following constraint at the transmitter:
\begin{itemize}
\item \cite[Eq. (8)]{Fabregas2010}: The bits $B_i$ in positions $i$ of the channel input are stochastically independent, i.e., the channel input pmf $\vecp$ is given by
\begin{align}
\vecp = \vecp^1\otimes\dotsb\otimes\vecp^m\label{eq:restrictionTransmitter}
\end{align}
where $\vecp^i$ is the pmf of $B_i$ and where $\otimes$ denotes the Kronecker product, see \cite[Def. 4.2.1]{Horn1991}.
\end{itemize}
According to \cite[Theorem~1]{martinez2009bit}, the following sum of mutual informations is an achievable rate for a BICM transceiver:
\begin{align}
\bicmmi(\vecp^1,\dotsc,\vecp^m):=\sum_{i=1}^m\miop(B_i;Y).
\end{align}
Following \cite[Eq. (19)]{Fabregas2010}, the ``BICM capacity'' $\capacity^\mathrm{bicm}$ is now given by
\begin{align}
\begin{split}
\capacity^\mathrm{bicm}=\max_{\vecp^1,\dotsc,\vecp^m}\quad&\bicmmi(\vecp^1,\dotsc,\vecp^m).
\end{split}
\end{align}
Unfortunately, the maximization is a non-convex problem. This will become clear in Sec.~\ref{sec:mibicm}.

\subsection{Problem Statement}
So far, BICM capacity has been calculated in literature via exhaustive search \cite{Agrell10b,Alvarado2011,Fabregas2010}. To determine the optimal bit pmfs with a precision of $\pm d$, $\bicmmi$ has to be evaluated $(\frac{1}{d})^m$ times, so the complexity of this approach increases exponentially in the number of bit positions $m$ and polynomially in the precision $d$. The objective of this work is to develop an algorithm that efficiently (compared to exhaustive search) calculates BICM capacity.

\section{Preliminary: $\bicmmi$ as a Function of $\pmb{p}^i$}
\label{sec:mibicm}

The goal of this section is to characterize the objective $\bicmmi$ as a function of one bit pmf $\vecp^i$. By this characterization, it will become clear that $\bicmmi$ is a non-convex function, and furthermore, we will see how we can maximize over $\vecp^i$. To this end, we pick an arbitrary bit position $i$ and assume that for each $j\neq i$, $B_j$ is distributed according to a fixed pmf and that $B_i$ is distributed according to a pmf that we interpret as a variable. To emphasize this distinction, we denote the pmfs for $j\neq i$ by $\hat{\vecp}^j$ and the pmf of $B_i$ by $\vecp^i$. The function $\bicmmi$ can now be written as
\begin{align}
\bicmmi(\vecp^1,\dotsc,\vecp^m)&=\sum_{j=1}^m[\entop(Y)-\entop(Y|B_j)]\\
&\hspace{-1cm}=m\entop(Y)-\entop(Y|B_i)-\sum_{j\neq i}\entop(Y|B_j).
\end{align}
We see that there are three kinds of terms that we need to express as functions of $\vecp^i$: the output entropy 
$\entop(Y)$, the conditional entropy $\entop(Y|B_i)$, and the conditional entropy $\entop(Y|B_j)$ for $j\neq i$.
\subsection{Output entropy as a function of $\pmb{p}^i$}
Define
\begin{align}
\vecq^i_0&:=\hat{\vecp}^1\otimes\dotsb\otimes\hat{\vecp}^{i-1}\otimes\begin{pmatrix}1&0\end{pmatrix}^T\otimes\hat{\vecp}^{i+1}\otimes\dotsb\otimes\hat{\vecp}^m\\
\vecq^i_1&:=\hat{\vecp}^1\otimes\dotsb\otimes\hat{\vecp}^{i-1}\otimes\begin{pmatrix}0&1\end{pmatrix}^T\otimes\hat{\vecp}^{i+1}\otimes\dotsb\otimes\hat{\vecp}^m.
\end{align}
The channel seen by the $i$th bit is now given by
\begin{align}
\mathh^i = \mathh\begin{pmatrix}\vecq^i_0&\vecq^i_1\end{pmatrix}\in\setr^{n\times 2}.
\end{align}
The output pmf can now be written as
\begin{align}
\vecr=\mathh\vecp=\mathh^i\vecp^i.
\end{align}
Thus, the output entropy as a function of $\vecp^i$ is given by
\begin{align}
\entop(Y)&=-\sum_{k=1}^n r_k\log r_k\\
&=-\sum_{k=1}^n (\vecr)_k\log (\vecr)_k\\
&=-\sum_{k=1}^n (\mathh^i\vecp^i)_k\log (\mathh^i\vecp^i)_k
\end{align}
where $(\vecx)_k$ denotes the $k$th entry of the vector $\vecx$. Since $-x\log x$ is concave in $x$, we conclude that the output entropy is concave in $\vecp^i$.
\subsection{Conditional entropy $\entop(Y|B_i)$ as a function of $\pmb{p}^i$}
The output entropy conditioned on the $i$th bit can be written as
\begin{align}
\entop(Y|B_i)=-\sum_{b=\texttt{0}}^\texttt{1} p^i_b\sum_{k=1}^n (\mathh^i)_{kb}\log(\mathh^i)_{kb}\label{eq:hybi}
\end{align}
where we index the rows of $\mathh^i$ by $1,\dotsc,k$ and the columns by the binary values $\texttt{0,1}$, e.g., $(\mathh^i)_{1\texttt{0}}$ is the entry of $\mathh^i$ in the first row and first column. 
 We conclude from \eqref{eq:hybi} that $\entop(Y|B_i)$ [and thereby $-\entop(Y|B_i)$, which contributes to the objective function] is linear in $\vecp^i$.
\subsection{Conditional entropy $\entop(Y|B_j)$ as a function of $\pmb{p}^i$}
Define
\begin{align}
\vecq^{ji}_{00}&:=\hat{\vecp}^1\otimes\dotsb\otimes\hat{\vecp}^{j-1}\otimes\begin{pmatrix}1&0\end{pmatrix}^T\otimes\hat{\vecp}^{j+1}\otimes\dotsb\nonumber\\
&\hspace{1.5cm}\otimes\hat{\vecp}^{i-1}\otimes\begin{pmatrix}1&0\end{pmatrix}^T\otimes\hat{\vecp}^{i+1}\otimes\dotsb\otimes\hat{\vecp}^m\\
\vecq^{ji}_{01}&:=\hat{\vecp}^1\otimes\dotsb\otimes\hat{\vecp}^{j-1}\otimes\begin{pmatrix}1&0\end{pmatrix}^T\otimes\hat{\vecp}^{j+1}\otimes\dotsb\nonumber\\
&\hspace{1.5cm}\otimes\hat{\vecp}^{i-1}\otimes\begin{pmatrix}0&1\end{pmatrix}^T\otimes\hat{\vecp}^{i+1}\otimes\dotsb\otimes\hat{\vecp}^m\\
\vecq^{ji}_{10}&:=\hat{\vecp}^1\otimes\dotsb\otimes\hat{\vecp}^{j-1}\otimes\begin{pmatrix}0&1\end{pmatrix}^T\otimes\hat{\vecp}^{j+1}\otimes\dotsb\nonumber\\
&\hspace{1.5cm}\otimes\hat{\vecp}^{i-1}\otimes\begin{pmatrix}1&0\end{pmatrix}^T\otimes\hat{\vecp}^{i+1}\otimes\dotsb\otimes\hat{\vecp}^m\\
\vecq^{ji}_{11}&:=\hat{\vecp}^1\otimes\dotsb\otimes\hat{\vecp}^{j-1}\otimes\begin{pmatrix}0&1\end{pmatrix}^T\otimes\hat{\vecp}^{j+1}\otimes\dotsb\nonumber\\
&\hspace{1.5cm}\otimes\hat{\vecp}^{i-1}\otimes\begin{pmatrix}0&1\end{pmatrix}^T\otimes\hat{\vecp}^{i+1}\otimes\dotsb\otimes\hat{\vecp}^m.
\end{align}
Now, the channel seen by the $j$th and the $i$th bit is given by
\begin{align}
\mathh^{ji}=\mathh\begin{pmatrix}\vecq^{ji}_{00}&\vecq^{ji}_{01}&\vecq^{ji}_{10}&\vecq^{ji}_{11}\end{pmatrix}\in\setr^{n\times 4}.
\end{align}
The channel seen by the $j$th bit can be written as
\begin{align}
\mathh^j=\mathh^{ji}\begin{pmatrix}
\vecp^i&\veczero\\
\veczero&\vecp^i
\end{pmatrix}.
\end{align}
Thus, the output entropy conditioned on the $j$th is
\begin{align}
&\entop(Y|B_j)=-\sum_{b=\texttt{0}}^\texttt{1} p^j_b\sum_{k=1}^n (\mathh^j)_{kb}\log(\mathh^j)_{kb}\\
&=-\sum_{b=\texttt{0}}^\texttt{1} p^j_b\sum_{k=1}^n \left[\mathh^{ji}\begin{pmatrix}
\vecp^i&\veczero\\
\veczero&\vecp^i
\end{pmatrix}\right]_{kb}\log\left[\mathh^{ji}\begin{pmatrix}
\vecp^i&\veczero\\
\veczero&\vecp^i
\end{pmatrix}\right]_{kb}.\label{eq:entropyji}
\end{align}
Since $-x\log x$ is concave in $x$, we conclude that $\entop(Y|B_j)$ is concave in $\vecp^i$. As a consequence, the term $-\entop(Y|B_j)$, which contributes to the objective function, is convex in $\vecp^i$. 
\subsection{Summary}
The objective function as a function of $\vecp^i$ can be characterized as follows:
\begin{align}
&\bicmmi(\hat{\vecp}^1,\dotsc,\hat{\vecp}^{i-1},\vecp^i,\hat{\vecp}^{i+1},\dotsc,\hat{\vecp}^m)\nonumber\\
&\hspace{1cm}=\underbrace{m\entop(Y)}_{\text{concave in }\vecp^i}-\underbrace{\entop(Y|B_i)}_{\text{linear in }\vecp^i}+\sum_{j\neq i}\underbrace{[-\entop(Y|B_j)]}_{\text{convex in }\vecp^i}.
\end{align}
As a sum of convex and concave terms, $\bicmmi$ is a non-convex function. However, as we detail in the next section, the convex-concave procedure \cite{Yuille2003} can be applied to maximize $\bicmmi$ over $\vecp^i$.

\section{\textsc{Bacm} Algorithm}
\label{sec:bacm}
\begin{figure}
\begin{algorithm}[(\textsc{Bacm})]\ 
\\
$\hat{\vecp}^1,\dotsc,\hat{\vecp}^m\leftarrow$ starting point\\
\textbf{repeat} \textcolor{blue}{\emph{bit alternation, outer loop}}\\
\indent\textbf{for} $i=1,\dotsc,m$ \textcolor{blue}{\emph{bit alternation, inner loop}}\\
\indent\indent maximize $\bicmmi$ over $\vecp^i$ \textcolor{blue}{\emph{see Alg.~\ref{alg:convexConcave}}}\\
\indent\indent update $\hat{\vecp}^i$ with the maximizing $\vecp^i$\\
\indent\textbf{end for}\\
\textbf{until} convergence
\label{alg:bacm}
\end{algorithm}
\end{figure}
\begin{figure}
\begin{algorithm}[(convex-concave procedure)]\ 
\\
calculate $\mathh^i$ and $\mathh^{ji}$, $j\neq i$\\
$\vecp^i\leftarrow\hat{\vecp}^i$ \\
\textbf{repeat}\\
\indent 1. $\hat{\vecp}^i\leftarrow \vecp^i$\\
\indent 2. $\displaystyle\vecp^i\leftarrow\argmax_{\vecp^i}f^i(\vecp^i,\hat{\vecp}^i)$ \textcolor{blue}{\emph{see Subsec.~\ref{subsec:inner}}}\\
\textbf{until} convergence
\label{alg:convexConcave}
\end{algorithm}
\vspace{-0.5cm}
\end{figure}
The objective $\bicmmi$ is a non-convex function of the pmfs $\vecp^1,\dotsc,\vecp^m$ with potentially more than one local maximum. Thus, finding an efficient algorithm that provably finds the global maximum is difficult. Therefore, we resort to the simpler problem of finding a local maximum. With a good starting point, the global maximum is nevertheless found by such an approach. To find local maxima, efficient methods are available. For the problem at hand, we choose the combination of two methods.
\begin{itemize}
\item We maximize over one bit pmf $\vecp^i$ at a time and then cycle through the $i=1,\dotsc,m$ until convergence. This approach goes under the name \emph{alternating maximization}.
\item To maximize over one bit pmf $\vecp^i$, we iteratively approximate $\bicmmi$ by a lower bound that is concave in $\vecp^i$ and maximize this concave lower bound. After convergence, the maximum of the concave lower bound is also a local maximum of $\bicmmi$ as a function of $\vecp^i$. This technique is known as the convex-concave procedure \cite{Yuille2003}.
\end{itemize}
We call this approach the \emph{bit-alternating convex-concave method} (\textsc{Bacm}). The alternating maximization over the bit pmfs is displayed in Alg.~\ref{alg:bacm}. The maximization over one bit pmf is detailed next.

\subsection{Concave Lower Bound}
As the objective is the sum of concave and convex functions, it cannot be maximized directly.
However, the \emph{convex-concave procedure} as defined in \cite[slide 26]{Boyd2008} can be applied. Define the function $h^j(\vecp^i)$ as the negative of the right-hand side of \eqref{eq:entropyji}. This function is convex in $\vecp^i$. The convex-concave procedure is an iterative method and works as follows. Denote by $\hat{\vecp}^i$ the result for $\vecp^i$ in the previous step. Then, in the current step, approximate $h^j(\vecp^i)$ by its first order Taylor expansion in $\hat{\vecp}^i$, i.e., by
\begin{align}
\hat{h}^j(\vecp^i,\hat{\vecp}^i):=h^j(\hat{\vecp}^i)+\nabla h^j(\hat{\vecp}^i)^T(\vecp^i-\hat{\vecp}^i).
\end{align}
Note that since $h^j(\vecp^i)$ is convex in $\vecp^i$ and the approximation $\hat{h}^j(\vecp^i,\hat{\vecp}^i)$ is linear in $\vecp^i$, the approximation $\hat{h}^j(\vecp^i,\hat{\vecp}^i)$ lower bounds $h^j(\vecp^i)$ for any value of $\vecp^i$. By a calculation similar to \cite[(7.61)--(7.63)]{Bocherer2012} it can be shown that $\hat{h}^j$ is given by
\begin{align}
&\hat{h}^j(\vecp^i,\hat{\vecp}^i)=\nonumber\\
&\sum_{b=\texttt{0}}^\texttt{1} p^j_b\sum_{k=1}^n \left[\mathh^{ji}\begin{pmatrix}
\vecp^i&\veczero\\
\veczero&\vecp^i
\end{pmatrix}\right]_{kb}\log\left[\mathh^{ji}\begin{pmatrix}
\hat{\vecp}^i&\veczero\\
\veczero&\hat{\vecp}^i
\end{pmatrix}\right]_{kb}.
\end{align}
Putting all together, we have a concave lower bound of $\bicmmi$ as a function of $\vecp^i$ given by
\begin{align}
&f^i(\vecp^i,\hat{\vecp}^i):=m\entop(Y)-\entop(Y|B_i)+\sum_{j\neq i}\hat{h}^j(\vecp^i,\hat{\vecp}^i)\label{eq:f0iapprox}\\
&=\!-m\sum_{k=1}^n (\mathh^i\vecp^i)_k\log (\mathh^i\vecp^i)_k\!+\!\sum_{b=\texttt{0}}^\texttt{1} p^i_b\sum_{k=1}^n (\mathh^i)_{kb}\log(\mathh^i)_{kb}\nonumber\\
& + \sum_{j\neq i}\sum_{b=\texttt{0}}^\texttt{1} p^j_b\sum_{k=1}^n \left[\mathh^{ji}\begin{pmatrix}
\vecp^i&\veczero\\
\veczero&\vecp^i
\end{pmatrix}\right]_{kb}\log\left[\mathh^{ji}\begin{pmatrix}
\hat{\vecp}^i&\veczero\\
\veczero&\hat{\vecp}^i
\end{pmatrix}\right]_{kb}\!\!.\label{eq:lowerBound}
\end{align}
Since $f^i$ is a concave function of $\vecp^i$, it can be maximized efficiently over $\vecp^i$, as we will explain in detail in the next subsection. We iteratively update $\hat{\vecp}^i$ with the value of $\vecp^i$ that maximizes $f^i(\vecp^i,\hat{\vecp}^i)$. Algorithm~\ref{alg:convexConcave} illustrates this procedure. After convergence, the pmf $\vecp^i$ locally maximizes $\bicmmi$ over $\vecp^i$ given the fixed pmfs $\hat{\vecp}^j$ for $j\neq i$.
\subsection{Solving the Inner Optimization Problem}
\label{subsec:inner}
We need to solve the optimization problem
\begin{align}
\maximize_{\text{pmf }\vecp^i}f^i(\vecp^i,\hat{\vecp}^i).\label{prob:pi}
\end{align}
Any pmf $\vecp^i$ can for some $p^i_0\in[0,1]$ be written as $\vecp^i=\left[\begin{smallmatrix}p^i_0\\1-p^i_0\end{smallmatrix}\right]$. We define
\begin{align}
f^i_0(p^i_0,\hat{\vecp}^i)=f^i(\left[\begin{smallmatrix}p^i_0\\1-p^i_0\end{smallmatrix}\right],\hat{\vecp}^i).
\end{align}
We can now formulate our optimization problem as
\begin{align}
\maximize_{p^i_0\in[0,1]}f^i_0(p^i_0,\hat{\vecp}^i).\label{prob:p0}
\end{align}
Note that the problems \eqref{prob:pi} and \eqref{prob:p0} are equivalent and furthermore, by \cite[Sec.~3.2.2]{Boyd2004}, $f^i_0$ is a concave function of $p_0$. Thus, our problem reduces to finding the maximum of a concave function with a scalar argument. This can be done as follows.

The first derivative of $H(Y)$, $H(Y|B_i)$, and $\hat{h}^j(\left[\begin{smallmatrix}p^i_0\\1-p^i_0\end{smallmatrix}\right],\hat{\vecp}^i)$, $j\neq i$ with respect to $p^i_0$ are respectively given by
\begin{align}
&\frac{\partial H(Y)}{\partial p^i_0}=-\sum_{k=1}^n\left\{\left[\mathh^i\bpm 1&-1\epm^T\right]_k\log(\mathh^i\left[\begin{smallmatrix}p^i_0\\1-p^i_0\end{smallmatrix}\right])_k\right.\nonumber\\
&\hspace{2cm}+\left.\left[\mathh^i\bpm 1&-1\epm^T\right]_k\right\}\\
&\frac{\partial H(Y|B_i)}{\partial p^i_0}=\sum_{k=1}^n\left[ (\mathh^i)_{k\texttt{0}}\log (\mathh^i)_{k\texttt{0}}-(\mathh^i)_{k\texttt{1}}\log (\mathh^i)_{k\texttt{1}}\right]\\
&\frac{\partial \hat{h}^j(\left[\begin{smallmatrix}p^i_0\\1-p^i_0\end{smallmatrix}\right],\hat{\vecp}^i)}{\partial p^i_0}=\sum_{b=\texttt{0}}^\texttt{1} p^j_b\sum_{k=1}^n\left[(\mathh^{ji})_{kb\texttt{0}}-(\mathh^{ji})_{kb\texttt{1}}\right]\nonumber\\
&\hspace{6cm}\cdot\log(\mathh^j)_{kb}
\end{align}
where we index the rows of $\mathh^{ji}$ by $k=1,\dotsc,n$ and the columns by the binary expansion $b_jb_i=\texttt{00,01,10,11}$. E.g., $(\mathh^{ji})_{1\texttt{10}}$ denotes the entry of $\mathh^{ji}$ in the $1$st row and the $3$rd column. For notational convenience, we write
\begin{align}
df^i_0(p_0,\hat{\vecp}^i):=\frac{\partial f^i_0(p^i_0,\hat{\vecp}^i)}{\partial p^i_0}.
\end{align}
Putting the expressions above together according to \eqref{eq:f0iapprox}, we get the first derivative of $f^i_0$. Since $f^i_0$ is concave, $d f^i_0$ is monotonically decreasing in $p^i_0$. Consequently, we can maximize $f^i_0$ over $p^i_0\in[0,1]$ as follows.
\begin{align}
\argmax_{p^i_0}f^i_0(p^i_0,\hat{\vecp}^i)=\begin{cases}
0&df^i_0(0^+,\hat{\vecp}^i)<0\\
1&df^i_0(1^-,\hat{\vecp}^i)>0\\
p^i_0:df^i_0(p^i_0,\hat{\vecp}^i)=0&\text{otherwise}.
\end{cases}\label{eq:solutionInnerProblem}
\end{align}
In our implementation \cite{website:bacm}, we use the bisection method to find $p^i_0$ in the third case. See Sec.~\ref{sec:complexity} for details.

\section{Adding an average cost constraint}
\label{sec:power}
We discuss how \textsc{Bacm} can be used to calculate BICM capacity when the bit pmfs are subject to an average cost constraint. Suppose we have a cost vector $\vecw\in\setrpp^{2^m}$, where $\setrpp$ denotes the set of positive real numbers. Then, the symbol costs seen by the $i$th bit are given by
\begin{align}
\vecw^i=[\vecw^T\!\begin{pmatrix}\vecq^i_0&\vecq^i_1\end{pmatrix}]^T.
\end{align}
The average cost can now be included by adding a weighted version of the average cost ${\vecw^i}^T\vecp^i$ to $f^i$, i.e., the inner optimization problem in Alg.~\ref{alg:convexConcave} now becomes
\begin{align}
\maximize_{\vecp^i} [f^i(\vecp^i,\hat{\vecp}^i)-\lambda {\vecw^i}^T\vecp^i].\label{eq:powerScaling}
\end{align}
This simply adds another linear term and our algorithm works in exactly the same way as before. Denote by $\vecp^{i*}$ the optimal pmfs found by this modified version of \textsc{Bacm} for some $\lambda$. Consider the resulting cost
\begin{align}
E=\vecw^T\vecp^*
\end{align}
where $\vecp^*=\vecp^{1*}\otimes\dotsb\otimes\vecp^{m*}$. Then, it can be shown that the bit pmfs $\vecp^{1*},\dotsc,\vecp^{m*}$ solve the optimization problem
\begin{align}
\begin{split}
\maximize_{\vecp^1,\dotsc,\vecp^m}\quad&\miop^\mathrm{bicm}(\vecp^1,\dotsc,\vecp^m)\\
\st\quad&\vecw^T(\vecp^{1}\otimes\dotsb\otimes\vecp^{m})\leq E.
\end{split}\label{eq:powerConstraint}
\end{align}

\section{Application to PAM in AWGN}
\begin{figure}
\footnotesize
\def\svgwidth{\columnwidth}
\executeiffilenewer{figures/bicmmi-edited2.svg}{figures/bicmmi-edited2.pdf}%
{inkscape -z -D --file=figures/bicmmi-edited2.svg %
--export-pdf=figures/bicmmi-edited2.pdf --export-latex}%
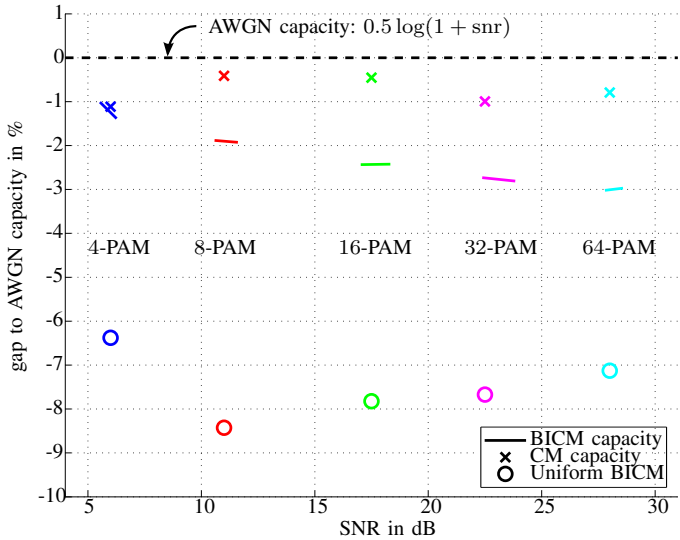%

\caption{Results for $4,8,16,32$, and $64$-PAM in AWGN. In the horizontal direction, SNR is displayed in dB. In the vertical direction, we show the gap in percent to the AWGN capacity $\capacity(\mathsf{snr})=0.5\log(1+\mathsf{snr})$. E.g., for BICM capacity, the gap is calculated as $100\cdot(1-\frac{\capacity(\mathsf{snr})}{\capacity^\mathrm{bicm}(\mathsf{snr})})$. For each constellation size and a corresponding target SNR, CM capacity, BICM capacity, and uniform BICM capacity are displayed. For BICM capacity, we display several values since we could adjust the effective SNR only via the weighting factor $\lambda$, see Sec.~\ref{sec:power}.}
\label{fig:awgn}
\vspace{-0.5cm}
\end{figure}
We use \textsc{Bacm} to calculate BICM capacity of PAM constellations in AWGN. To calculate the BICM capacity of PAM constellations in AWGN, optimization has to be done over the labeling of the signal points, the scaling of the constellation, and the bit pmfs, see \cite[Eq.~(40)]{Agrell10b} for details. Here, we fix the labeling to the binary reflected Gray code \cite[Sec.~II-B]{Agrell10b} and optimize over constellation scaling and bit pmfs. To be able to use \textsc{Bacm}, we discretize the channel output into $200$ equally spaced points. For each scaling, the discretized AWGN channel with $M=2^m$ constellation points at the input can thus be represented by a DMC specified by a transition matrix $\mathh\in\setr^{200\times M}$. For this DMC, we use the method proposed in Sec.~\ref{sec:power} to calculate the BICM capacity. To achieve a target SNR, we iteratively adapt the weighting $\lambda$ of the average power in \eqref{eq:powerScaling}. We repeat this for different constellation scalings and choose the scaling that yields the largest value for $\bicmmi$. This largest value is the BICM capacity and we denote it by $\capacity^\mathrm{bicm}(\mathsf{snr})$. Results for $4,8,16,32$, and $64$-PAM are displayed in Fig.~\ref{fig:awgn}. For comparison, coded modulation (CM) capacity \cite[Eq.~(28)]{Agrell10b} of the corresponding constellation and $\bicmmi$ for uniform bit pmfs are displayed. The values for CM capacity were obtained via CVX \cite{CVX}. The BICM capacity significantly outperforms uniform BICM and gets close to CM capacity. We calculated the optimal bit pmfs with a precision of $d=10^{-5}$.

\section{Complexity of \textsc{Bacm}}
\label{sec:complexity}
We start by analyzing the complexity of the inner optimization problem. To cover the first two cases in \eqref{eq:solutionInnerProblem}, we need to evaluate $df^i_0$ two times. To find the $p^i_0$ in the third case we use the bisection method starting with the upper bound $u=1$ and the lower bound $\ell=0$, and we terminate when $u-\ell\leq 2d$. After termination, we assign $p^i_0=\frac{u+\ell}{2}$. Thus, we calculate $p^i_0$ with a precision of $\pm d$. According to \cite[p.~146]{Boyd2004}, the number of times we need to evaluate $df^i_0$ until termination is given by
\begin{align}
\left\lceil\log_2\frac{u-\ell}{2d}\right\rceil&=\left\lceil\log_2\frac{1-0}{2d}\right\rceil=\left\lceil\log_2 \frac{1}{2d}\right\rceil.
\end{align}
When evaluating $df^i$, by \eqref{eq:f0iapprox}, we need to evaluate $\partial\hat{h}^j/\partial p^i_0$ for each $j\neq i$, which results in a number of $m-1$ or roughly $m$ evaluations. Overall, the number of evaluations needed for solving the inner optimization problem once is roughly $m\log_2\frac{1}{2d}$. The sizes of the matrices involved in $\eqref{eq:lowerBound}$ are invariant under $m$, i.e., $\mathh^{ji}\in\setr^{n\times 4}$ and $\mathh^i\in\setr^{n\times 2}$. Therefore, the number of iterations until convergence in Alg.~\ref{alg:convexConcave} should be approximately invariant under $m$ and we denote it by a constant $K$. For our AWGN simulations, this number was around $K=3$, independent of $m$. The complexity of maximizing $\bicmmi$ over one bit pmf is thus approximately $Km\log_2\frac{1}{2d}$. This maximization has to be done for $i=1,\dotsc,m$, i.e., $m$ times, which adds another factor of $m$ to the complexity. This procedure has to be repeated $L$ times until convergence in the outer loop of Alg.~\ref{alg:bacm}. This number depends on $m$. For the AWGN simulations, we observed for $m=2,3,4,5,6$, respectively, the values
\begin{align}
2.00\qquad3.27\qquad3.90\qquad4.24\qquad4.31.\label{eq:ba}
\end{align}
The average for each $m$ is taken separately over all values that were observed when executing $\textsc{Bacm}$. This value increases slightly with $m$. To have a rough bound on complexity, we assume that $L$ increases at most linearly with $m$, which is consistent with the observed data \eqref{eq:ba}. All together, we have a complexity that is approximately of the order $LKm^2\log_2\frac{1}{2d}\leq Km^3\log_2\frac{1}{2d}$. In summary, \textsc{Bacm} scales as $m^3$ and logarithmically in the precision $d$.

\bibliographystyle{IEEEtran}
\bibliography{IEEEabrv,confs-jrnls,bacm-references,references_all_AA}

\end{document}